\pdfoutput=1
\documentclass[conference]{IEEEtran}
\IEEEoverridecommandlockouts
\usepackage{cite}
\usepackage{amsmath,amssymb,amsfonts}
\usepackage{algorithmic}
\usepackage{graphicx}
\usepackage{textcomp}
\usepackage{xcolor}

\def\BibTeX{{\rm B\kern-.05em{\sc i\kern-.025em b}\kern-.08em
    T\kern-.1667em\lower.7ex\hbox{E}\kern-.125emX}}

\usepackage[absolute]{textpos}

\begin{document}

\IEEEoverridecommandlockouts
\IEEEpubid{\makebox[\columnwidth]{ 979-8-3315-6337-0/26/\$31.00 \copyright2026 IEEE \hfill} \hspace{\columnsep}\makebox[\columnwidth]{ }}

\begin{textblock}{5}(12.5,0.8)
\end{textblock}

\title{Can Agents Secure Hardware? Evaluating Agentic LLM-Driven Obfuscation for IP Protection\\
\thanks{* Corresponding author: Sujan Ghimire (sghimire@arizona.edu)}
}

\author{
    \IEEEauthorblockN{Sujan Ghimire\textsuperscript{1 *},  Parsa Mirfasihi\textsuperscript{1}, Muhtasim Alam Chowdhury\textsuperscript{1}, 
    Veeramani Pugazhenthi\textsuperscript{1},\\
    Harish Kumar Dharavath\textsuperscript{1}, Farshad Firouzi\textsuperscript{4}, Rozhin Yasaei\textsuperscript{3}, Pratik Satam\textsuperscript{1,2}, Soheil Salehi\textsuperscript{1}}
    \IEEEauthorblockA{
        \textsuperscript{1}Department of Electrical and Computer Engineering, University of Arizona, Tucson, AZ, USA \\
        \textsuperscript{2}Department of Systems and Industrial Engineering, University of Arizona, Tucson, AZ, USA\\         \textsuperscript{3}College of Information Science, University of Arizona, Tucson, AZ, USA\\ 
        \textsuperscript{4}School of Electrical, Computer and Energy Engineering, Arizona State University, Tempe, AZ, USA\\
        \{\textsuperscript{1}sghimire, \textsuperscript{1}parsamirfasihi, \textsuperscript{1}mmc7, \textsuperscript{1}veerpugazh5, \textsuperscript{1}harrydhara16, \textsuperscript{3}yasaei, \textsuperscript{1,2}pratiksatam, \textsuperscript{1}ssalehi\}@arizona.edu, \\\textsuperscript{4}farshad.firouzi@asu.edu
        }
        }

\maketitle

\begin{abstract}
The globalization of integrated circuit (IC) design and manufacturing has increased the exposure of hardware intellectual property (IP) to untrusted stages of the supply chain, raising concerns about reverse engineering, piracy, tampering, and overbuilding. Hardware netlist obfuscation is a promising countermeasure, but automating the generation of functionally correct and security-relevant obfuscated circuits remains challenging, particularly for benchmark-scale designs. This paper presents an agentic, large language model (LLM)-driven framework for automated hardware netlist obfuscation. The proposed framework combines retrieval-grounded planning, structured lock-plan generation, deterministic netlist compilation, functional verification, and SAT-based security evaluation. Rather than a single prompt-to-output generation step, the framework decomposes the task into specialized stages for circuit analysis, synthesis, verification, and attack evaluation. We evaluate the framework on ISCAS-85 benchmarks using functional equivalence checking and SAT-based attacks. Results show that the framework generates correct locked netlists while introducing measurable output corruption under incorrect keys, while SAT attacks remain effective. These findings highlight both the potential and current limitations of agentic LLM-driven obfuscation.
\end{abstract}

\begin{IEEEkeywords}
Hardware Security, Logic Obfuscation, Large Language Model (LLM), Agentic AI, SAT attack.
\end{IEEEkeywords}

\section{Introduction}
The globalization of IC design and manufacturing has increased the exposure of hardware IP to untrusted stages of the supply chain, raising concerns about reverse engineering, piracy, tampering, overbuilding, and counterfeiting. IP obfuscation addresses this problem by inserting key-controlled logic into a design such that correct behavior is obtained only when the proper key is applied. Without the correct key, the circuit exhibits degraded functionality, making unauthorized reproduction more difficult.

Currently, LLMs have shown strong performance in code synthesis, planning, debugging, and domain-specific design automation. Recent work suggests that LLMs can assist with HDL generation and verification, motivating agentic AI systems in which models operate within multi-stage workflows that reason, invoke tools, and iteratively refine outputs. This paradigm is particularly relevant to IP obfuscation, where syntactic generation alone is insufficient. A generated obfuscated circuit must remain structurally valid, preserve correct-key functionality, induce corruption under incorrect keys, and withstand security evaluation. Prompt-only generation is therefore inadequate. The problem requires an agentic workflow capable of planning, generating, verifying, evaluating, and refining obfuscation schemes.

This paper presents an agentic, LLM-driven framework for automated IP obfuscation of hardware netlists. 
The framework combines retrieval-grounded planning, candidate lock synthesis, deterministic netlist compilation, functional verification, security-oriented evaluation, and iterative refinement. The planner selects lock targets and strategies, the synthesis stage generates structured lock plans, deterministic modules render valid obfuscated netlists, and verification and attack-evaluation stages assess correctness and security behavior. The framework is evaluated on benchmark circuits using metrics including parse validity, correct-key functionality, wrong-key corruption, runtime, and SAT-based security evaluation. The main contributions of this work are:


\begin{itemize}
\item An agentic LLM-driven framework for automated IP obfuscation integrating planning, synthesis, verification, and security evaluation.
\item A structured lock-plan generation approach for reliable LLM-guided hardware transformations.
\item A quantitative evaluation methodology using functional correctness and SAT-based analysis.
\item An empirical study on benchmark circuits demonstrating the feasibility and limitations of AI-assisted hardware obfuscation.
\end{itemize}

\section{BACKGROUND}
\subsection{LLMs in Hardware Design and Security}

Recent advances in LLMs have expanded their role beyond natural language processing to code generation, planning, debugging, and domain-specific design automation, making them relevant to hardware design and security workflows \cite{Ghimire2025_Survey, Latibari2025}. Prior work in hardware design spans design assistance, optimization, verification, and debugging. Chip-Chat demonstrated that general-purpose LLMs can support hardware design tasks such as microprocessor development~\cite{blocklove2023chip}. VeriGen improved HDL generation via fine-tuning on Verilog corpora~\cite{thakur2023verigen}, while VeriPPA and ChipGPT enhanced PPA optimization and design-space exploration~\cite{thorat2023advanced,changimproving}. In verification and debugging, approaches such as AssertLLM, ChiRAAG, RTLFixer, and HDLDebugger enable assertion generation, syntax repair, and retrieval-enhanced debugging~\cite{fang2024assertllm,mali2024chiraag,tsai2023rtlfixer,yao2024hdldebugger}. 
LLMs have also been applied to hardware security, including Trojan generation and detection (SENTAUR), offensive security analysis, and vulnerability assessment frameworks~\cite{bhandari2024sentaursecurityenhancedtrojan,kokolakis2024harnessing,Ghimire2025_Hwrex,Lin2025,Bandi2021,Hassan2023,Asmita2025}. Prior LLM-based IP obfuscation efforts demonstrate promise but remain limited for larger benchmark circuits~\cite{latibari2024automated}. More recently, research has shifted toward agentic, multi-stage workflows involving planning, tool use, and iterative refinement, enabling more reliable automation.

\subsection{IP Obfuscation and Security Evaluation}

IP obfuscation inserts key-controlled logic such that correct functionality is obtained only under the correct key, protecting designs against reverse engineering and piracy. Early approaches relied on simple key-gate insertion with limited resilience, motivating the development of structured, attack-aware obfuscation methods \cite{Gandhi2024,Gubbi2024}. Recent work explores optimization-driven and structurally resilient techniques, including learning-resilient obfuscation, corrupt-and-correct schemes, testability-aware methods, sub-circuit replacement, and hybrid strategies \cite{Wang2025,Aksoy2024,Pandi2025,Rathor2024}.

Evaluation has become increasingly rigorous: effective schemes must preserve correct-key functionality, induce corruption under wrong keys, and withstand adversarial analysis. Accordingly, prior work employs formal and quantitative evaluation using pseudo-Boolean analysis, SMT/SAT-based frameworks, and learning-based attacks \cite{Merten2023,Han2025,Ahmed2024,McDaniel2024}. These developments position IP obfuscation as a synthesis-and-validation problem requiring joint consideration of correctness, overhead, and attack resilience \cite{Gandhi2024}.

\subsection{Agentic LLM Workflows}

Recent advances in LLMs have shifted from single-prompt generation to agentic workflows. While single-shot prompting can produce outputs for small tasks, it struggles with complex, multi-step reasoning \cite{latibari2024automated}. Agentic workflows address this by enabling LLMs to decompose tasks, invoke tools, inspect intermediate outputs, and iteratively refine results. Previous works shows that modular task decomposition, workflow generation, and coordinated agents improve robustness in complex reasoning environments \cite{niu2025flowmodularizedagenticworkflow,Xiong2025SelfOrganizingAN}. 

Other studies emphasize context management, validation, and structured coordination for maintaining correctness \cite{Chang2025}. Hybrid frameworks combining LLM reasoning with formal methods further improve reliability and scalability \cite{Li2024,Gundawar2024}. In hardware domains, where structural and functional constraints are strict, retrieval-grounded generation and verification-aware reasoning reduce hallucinations and improve correctness \cite{Ayalasomayajula2024,Quddus2024}. 

These works highlight a gap at the intersection of hardware security and agentic AI. While obfuscation research has advanced attack-aware methods and evaluation, and LLM research has enabled multi-stage workflows, limited work combines these directions into a unified framework for circuit-conditioned logic-lock synthesis. This gap motivates the proposed framework, which integrates planning, lock synthesis, deterministic netlist generation, functional verification, and SAT-based security evaluation.

\section{Agentic IP obfuscation Framework}
\label{sec:framework}

The proposed framework is an agentic, LLM-driven IP obfuscation pipeline that converts an input \texttt{.bench} circuit into an obfuscated candidate through planning, synthesis, verification, security evaluation, and refinement. Unlike one-shot prompt-based generation, the workflow separates reasoning from validation. LLM agents perform planning, candidate generation, and refinement, while deterministic modules handle parsing, netlist rendering, functional verification, and attack evaluation.

\begin{figure}[t]
    \centering
    \includegraphics[width=0.96\linewidth]{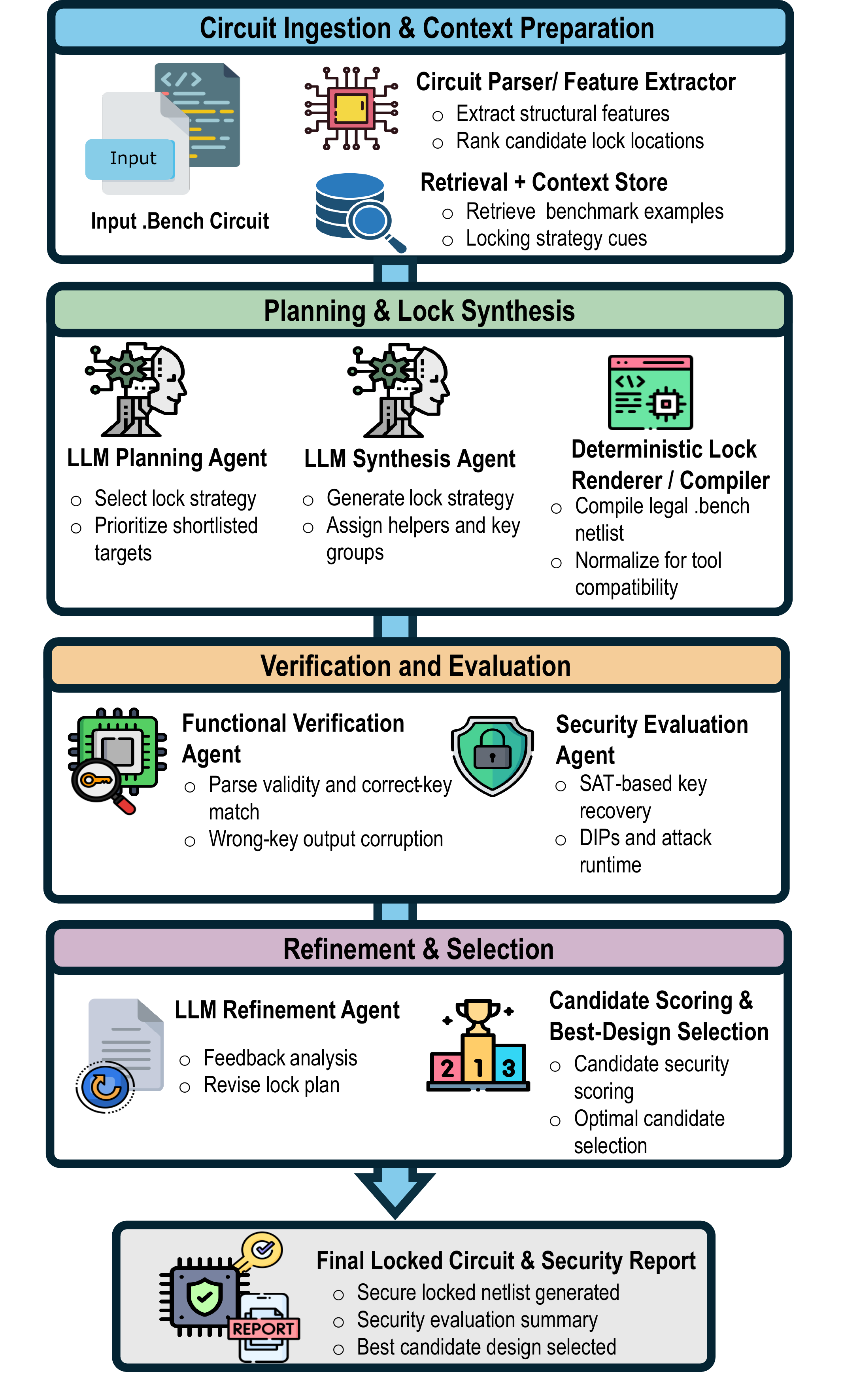}
    \caption{Overview of the proposed agentic LLM-driven IP obfuscation framework combining planning, synthesis, verification, SAT-based security evaluation, and refinement.}
    \label{Fig:framwwork}
\end{figure}

As illustrated in Fig.~\ref{Fig:framwwork}, the flow starts by parsing the input circuit and extracting structural features such as gate count, depth, fanout, and output-cone information. These features rank candidate lock locations using a topology-aware heuristic motivated by prior logic-cone and testability-driven approaches \cite{Rajendran2012, Lee2015, Pandi2025}. The framework prioritizes nodes with stronger downstream influence, depth, observability, and output-cone coverage, since perturbations at such locations are more likely to propagate to outputs. The resulting ranked list is used to construct a shortlist of candidate targets. In parallel, a retrieval agent collects benchmark examples and hardware-security context, and the planning agent selects an obfuscation style and target set.
The synthesis agent returns a structured lock plan specifying selected targets, lock style, helper signals, and key-bit groupings, which is compiled into a valid obfuscated netlist. This intermediate representation reduces syntax errors, constrains valid signal usage, and enables deterministic compilation. The renderer normalizes the circuit into a tool-compatible gate basis for ABC and Yosys. The implementation supports \texttt{xor\_xnor}, \texttt{perturb\_restore}, \texttt{mux\_lock}, \texttt{pairwise\_subgraph}, and \texttt{hybrid} styles, and multiple candidates can be generated per circuit.

Each candidate undergoes functional verification to check parse validity, correct-key behavior, and wrong-key corruption. Correct-key consistency is evaluated through simulation against the original circuit, while wrong-key corruption is measured as output mismatch under incorrect keys. Structural overhead metrics such as gate overhead and key-input count are additionally reported.

Candidates are then evaluated using SAT-based or enumeration-based key recovery. A PySAT-based distinguishing-input-pattern attack iteratively generates oracle-guided constraints until the key space collapses or a DIP budget is reached. The attack reports key recovery, distinguishing inputs, runtime, and remaining key space, providing a quantitative security assessment. If a candidate fails parsing, shows weak corruption, or is easily recovered, a refinement agent uses verification and security feedback to generate an improved lock plan, which is recompiled and reevaluated.

Finally, candidates are ranked using a heuristic that rewards correct-key functionality and higher corruption while penalizing invalid outputs and excessive overhead. The best candidate is selected as the final obfuscated netlist, enabling a closed-loop workflow that combines LLM-driven decision-making with deterministic verification and security evaluation.
\section{Experimental Setup and Results}
\label{sec:setup_results}

\subsection{Experimental Setup}

The framework is evaluated using ISCAS-85 combinational circuits in \texttt{.bench} format~\cite{iscas85}, including \texttt{c432}, \texttt{c499}, \texttt{c880}, \texttt{c1355}, \texttt{c1908}, \texttt{c3540}, \texttt{c5315}, and \texttt{c7552}. These benchmarks span a range of structural complexity and are widely used in hardware security research. Sequential circuits are not considered.

For each benchmark, the framework generates obfuscated circuits using the workflow in Section~\ref{sec:framework}. We evaluate three models: \texttt{gpt-5}, \texttt{llama3.1:8b}, and \texttt{qwen2.5-coder:14b}, all using the same structured lock-plan interface. Experiments use 8-, 16-, and 32-bit keys with one candidate per benchmark--key pair and SAT-based evaluation.

Each design is evaluated using four metric groups: \textit{functional validity} (parse success, correct-key match, wrong-key corruption), \textit{structural cost} (gate overhead, key-gate count, key-input count), \textit{security metrics} (SAT success, runtime, DIPs, remaining keys), and \textit{workflow metrics} (runtime and LLM usage).

An external equivalence check is performed by translating \texttt{.bench} netlists to Verilog, binding the correct key, and verifying equivalence using Yosys. This serves as post-generation validation, while reported results rely on simulation-based correctness within the pipeline.

\subsection{Results}

All evaluated models generate syntactically valid obfuscated netlists that preserve correct-key functionality, demonstrating that the framework reliably produces usable designs. We evaluate (1) wrong-key corruption, (2) SAT-based security, and (3) cross-model behavior.

\subsubsection{Functional Validity and Corruption Behavior}

\begin{figure}[t!]
\centering
\includegraphics[width=\columnwidth]{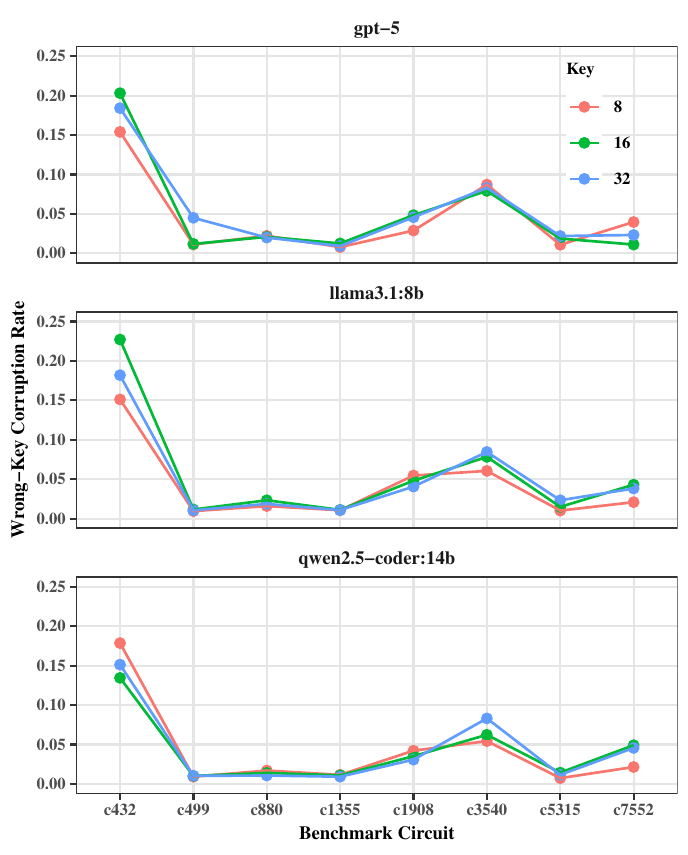}
\caption{Wrong-key corruption rate across ISCAS-85 benchmarks for GPT-5, LLaMA-3.1-8B, and Qwen-2.5-Coder-14B with 8-, 16-, and 32-bit keys.}
\label{fig:result1}
\end{figure}

Figure~\ref{fig:result1} shows wrong-key corruption across benchmarks. All runs preserve correct-key functionality while producing non-zero corruption under incorrect keys. Corruption varies by circuit, ranging from approximately 0.01 to 0.23, with \texttt{c432} showing the highest and \texttt{c5315} the lowest values. This indicates strong dependence on circuit topology and signal propagation.

Increasing key size does not consistently increase corruption, suggesting that lock placement and structural influence dominate over key length. We next evaluate resistance to SAT-based attacks.
 
\subsubsection{Security Evaluation}

\begin{figure}[t!]
\centering
\includegraphics[width=0.9\columnwidth]{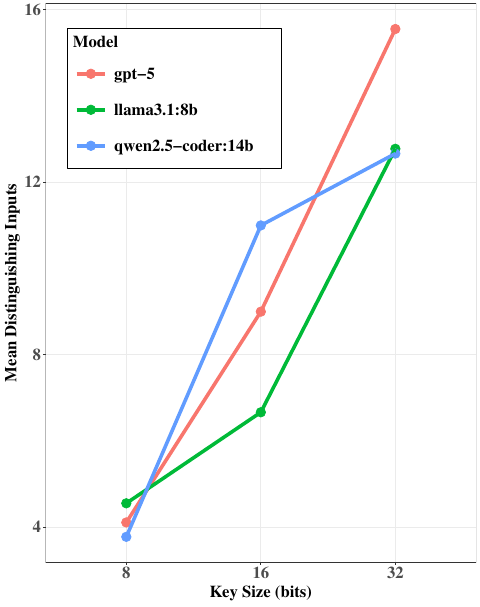}
\caption{Mean distinguishing input patterns (DIPs) required by SAT attacks for different key sizes.}
\label{fig:result2}
\end{figure}
															
Figure~\ref{fig:result2} reports the mean number of DIPs required for key recovery. DIPs increase with key size across all models, indicating higher attack effort. However, the SAT solver successfully recovers the correct key in all cases, showing that current obfuscation templates increase attack cost but do not prevent recovery.

Attack effort also varies across benchmarks, indicating that circuit structure and lock placement significantly influence recoverability. We next compare model behavior.

\subsubsection{Model Comparison}

All models generate valid obfuscated circuits with correct-key functionality and measurable corruption, demonstrating consistent performance across proprietary and open-source LLMs. Corruption and SAT effort vary across circuits, with smaller circuits such as \texttt{c432} showing higher corruption and larger circuits such as \texttt{c5315} and \texttt{c7552} showing lower values. Key size increases attack effort but not uniformly.

GPT-5 generally shows higher SAT effort but higher runtime. \texttt{LLaMA-3.1-8B} achieves comparable corruption with lower runtime, while \texttt{Qwen-2.5-Coder-14B} shows intermediate behavior. Table~\ref{tab:model_compare} summarizes aggregate results. Overall, the framework consistently generates valid obfuscated circuits with measurable corruption, while SAT evaluation shows increased attack effort with key size but full key recovery in all cases. These results highlight both the effectiveness of the framework and the limitations of current obfuscation strategies.

\begin{table}[t]
\centering
\caption{Aggregate statistics from the experimental dataset (72 runs).}
\label{tab:model_compare}
\small
\begin{tabular}{lccc}
\hline
\textbf{Model} & \textbf{Corr.} & \textbf{DIPs} & \textbf{Runtime (s)} \\
\hline
GPT-5              & 0.050 & 9.6 & 209 \\
LLaMA-3.1-8B       & 0.051 & 8.0 & 66 \\
Qwen-2.5-Coder-14B & 0.045 & 9.1 & 114 \\
\hline
\end{tabular}
\end{table}

\section{Discussion}

The results show that the proposed framework can reliably generate functionally correct obfuscated circuits with measurable wrong-key corruption across a range of benchmarks and models. However, the consistently successful SAT-based key recovery indicates that current lock templates increase attack effort but do not provide strong resistance against modern attacks. Moreover, obfuscation does not eliminate vulnerability to physical side-channel attacks, where power leakage can expose internal computation behavior without requiring key recovery \cite{Veeramani2025}. The observed variation in corruption and attack difficulty across circuits highlights the importance of topology-aware lock placement, as structural characteristics such as logic depth and output cone affect both corruption propagation and attack resilience. These findings suggest that effective obfuscation requires tighter integration between structural analysis and security objectives, highlighting the limitations of existing strategies under strong adversarial models.
   
\section{Conclusion}    
This paper presented an agentic, LLM-driven framework for automated hardware IP obfuscation that integrates planning, structured lock synthesis, deterministic netlist generation, functional verification, and SAT-based security evaluation. The results show that the framework consistently generates valid obfuscated circuits with correct-key functionality and measurable wrong-key corruption. However, SAT-based analysis reveals that existing obfuscation templates remain vulnerable to key recovery, even as attack effort increases with key size. These findings highlight both the promise of LLM-assisted obfuscation and the need for more robust, security-aware design strategies.


\bibliographystyle{ieeetr}
\bibliography{ref}

@misc{iscas85,
  title = {{ISCAS} '85 Benchmarks - Verilog},
  note = {Accessed October 05, 2024},
}

@article{blocklove2023chip,
  title={{Chip-Chat: Challenges and Opportunities in Conversational Hardware Design}},
  author={Blocklove, Jason and Garg, Siddharth and Karri, Ramesh and Pearce, Hammond},
  journal={arXiv preprint arXiv:2305.13243},
  year={2023}
}

@misc{thakur2023verigen,
      title={{VeriGen: A Large Language Model for Verilog Code Generation}}, 
      author={Shailja Thakur and Baleegh Ahmad and Hammond Pearce and Benjamin Tan and Brendan Dolan-Gavitt and Ramesh Karri and Siddharth Garg},
      year={2023},
      eprint={2308.00708},
      archivePrefix={arXiv},
      primaryClass={cs.PL}
}

@inproceedings{latibari2024automated,
  title={{Automated Hardware Logic Obfuscation Framework Using GPT}},
  author={Latibari, Banafsheh Saber and Ghimire, Sujan and Chowdhury, Muhtasim Alam and Nazari, Najmeh and Gubbi, Kevin Immanuel and Homayoun, Houman and Sasan, Avesta and Salehi, Soheil},
  booktitle={2024 IEEE 17th Dallas Circuits and Systems Conference (DCAS)},
  pages={1--5},
  year={2024},
  organization={IEEE}
}

@article{thorat2023advanced,
  title={{Advanced Language Model-Driven Verilog Development: Enhancing Power, Performance, and Area Optimization in Code Synthesis}},
  author={Thorat, Kiran and Zhao, Jiahui and Liu, Yaotian and Peng, Hongwu and Xie, Xi and Lei, Bin and Zhang, Jeff and Ding, Caiwen},
  journal={arXiv preprint arXiv:2312.01022},
  year={2023}
}

@article{changimproving,
  title={{Improving Large Language Model Hardware Generating Quality through Post-LLM Search}},
  author={Chang, Kaiyan and Ren, Haimeng and Wang, Mengdi and Liang, Shengwen and Han, Yinhe and Li, Huawei and Li, Xiaowei and Wang, Ying}
}

@misc{fang2024assertllm,
      title={{AssertLLM: Generating and Evaluating Hardware Verification Assertions from Design Specifications via Multi-LLMs}}, 
      author={Wenji Fang and Mengming Li and Min Li and Zhiyuan Yan and Shang Liu and Hongce Zhang and Zhiyao Xie},
      year={2024},
      eprint={2402.00386},
      archivePrefix={arXiv},
      primaryClass={cs.AR}
}

@misc{mali2024chiraag,
      title={{ChIRAAG: ChatGPT Informed Rapid and Automated Assertion Generation}}, 
      author={Bhabesh Mali and Karthik Maddala and Sweeya Reddy and Vatsal Gupta and Chandan Karfa and Ramesh Karri},
      year={2024},
      eprint={2402.00093},
      archivePrefix={arXiv},
      primaryClass={cs.SE}
}

@article{tsai2023rtlfixer,
  title={{RTLFixer: Automatically Fixing RTL Syntax Errors with Large Language Models}},
  author={Tsai, YunDa and Liu, Mingjie and Ren, Haoxing},
  journal={arXiv preprint arXiv:2311.16543},
  year={2023}
}

@article{yao2024hdldebugger,
  title={{HDLdebugger: Streamlining HDL debugging with Large Language Models}},
  author={Yao, Xufeng },
  journal={arXiv preprint arXiv:2403.11671},
  year={2024}
}

@misc{bhandari2024sentaursecurityenhancedtrojan,
      title={{SENTAUR: Security EnhaNced Trojan Assessment Using LLMs Against Undesirable Revisions}}, 
      author={Jitendra Bhandari and Rajat Sadhukhan and Prashanth Krishnamurthy and Farshad Khorrami and Ramesh Karri},
      year={2024},
      eprint={2407.12352},
      archivePrefix={arXiv},
      primaryClass={cs.CR}, 
}

@inproceedings{kokolakis2024harnessing,
  title={{Harnessing the Power of General-Purpose LLMs in Hardware Trojan Design}},
  author={Kokolakis, Georgios and Moschos, Athanasios and Keromytis, Angelos D},
  booktitle={International Conference on Applied Cryptography and Network Security},
  pages={176--194},
  year={2024},
  organization={Springer}
}

@INPROCEEDINGS{Gandhi2024,
  author={Gandhi, Jugal and Shekhawat, Diksha and Santosh, M. and Pandey, Jai Gopal},
  booktitle={2024 IEEE 17th International Symposium on Embedded Multicore/Many-core Systems-on-Chip (MCSoC)}, 
  title={{Emerging Frontiers and Limitations of Logic Locking for Secure IC Design}}, 
  year={2024},
  volume={},
  number={},
  pages={239-244},
  doi={10.1109/MCSoC64144.2024.00047}}

@article{Wang2025,
   author = {Zeng Wang and Lilas Alrahis and Animesh Basak Chowdhury and Dominik Germek and Ramesh Karri and Ozgur Sinanoglu},
   doi = {10.1109/ACCESS.2025.3612444},
   issn = {21693536},
   journal = {IEEE Access},
   pages = {166649-166669},
   publisher = {Institute of Electrical and Electronics Engineers Inc.},
   title = {{OptiLock: Automated Optimization of Learning-Resilient Logic Locking}},
   volume = {13},
   year = {2025}
}

@article{Pandi2025,
   author = {Marziye Pandi and Mostafa Moghaddas and Hakem Beitollahi},
   doi = {10.1007/s11227-025-07803-9},
   isbn = {0123456789},
   issn = {15730484},
   issue = {14},
   journal = {The Journal of Supercomputing 2025 81:14},
   month = {9},
   pages = {1320-},
   publisher = {Springer},
   title = {{TestLock: A Testability Logic Locking Method Against Machine Learning-based Oracle-Less Attacks}},
   volume = {81},
   year = {2025}
}

@article{Aksoy2024,
   author = {Levent Aksoy and Muhammad Yasin and Samuel Pagliarini},
   doi = {10.1109/LATS62223.2024.10534592},
   isbn = {9798350365559},
   journal = {2024 IEEE 25th Latin American Test Symposium, LATS 2024},
   publisher = {Institute of Electrical and Electronics Engineers Inc.},
   title = {{CAC 2.0: A Corrupt and Correct Logic Locking Technique Resilient to Structural Analysis Attacks}},
   year = {2024}
}

@article{Rathor2024,
   author = {Vijaypal Singh Rathor and Munesh Singh and Kshira Sagar Sahoo and Saraju P. Mohanty},
   month = {6},
   title = {{SubLock: Sub-Circuit Replacement based Input Dependent Key-based Logic Locking for Robust IP Protection}},
   year = {2024}
}

@article{Merten2023,
   author = {Marcel Merten and Muhammad Hassan and Rolf Drechsler},
   doi = {10.1109/DDECS57882.2023.10139590},
   isbn = {9798350332773},
   journal = {Proceedings - 2023 26th International Symposium on Design and Diagnostics of Electronic Circuits and Systems, DDECS 2023},
   pages = {105-110},
   publisher = {Institute of Electrical and Electronics Engineers Inc.},
   title = {{Quality Assessment of Logic Locking Mechanisms using Pseudo-Boolean Optimization Techniques}},
   year = {2023}
}

@article{Han2025,
   author = {Zhaokun Han and Daniel Xing and Kostas Amberiadis and Ankur Srivastava and Jeyavijayan Jv Rajendran},
   doi = {10.1109/DAC63849.2025.11132623},
   isbn = {9798331503048},
   issn = {0738100X},
   journal = {Proceedings - Design Automation Conference},
   publisher = {Institute of Electrical and Electronics Engineers Inc.},
   title = {{SCONE: A Logic Locking Technique Utilizing SMT Solver and Circuit Encoding Scheme for Efficient Hardware IP Protection}},
   year = {2025}
}

@article{Ahmed2024,
   author = {Bulbul Ahmed and Sazadur Rahman and Kimia Zamiri Azar and Farimah Farahmandi and Fahim Rahman and Mark Tehranipoor},
   doi = {10.1145/3649476.3660382},
   isbn = {9798400706059},
   journal = {Proceedings of the ACM Great Lakes Symposium on VLSI, GLSVLSI},
   month = {6},
   pages = {489-494},
   publisher = {Association for Computing Machinery},
   title = {{SeeMLess: Security Evaluation of Logic Locking using Machine Learning oriented Estimation}},
   year = {2024}
}

@article{McDaniel2024,
   author = {Isaac McDaniel and Michael Zuzak and Ankur Srivastava},
   doi = {10.1145/3674903},
   issn = {15577309},
   issue = {4},
   journal = {ACM Transactions on Design Automation of Electronic Systems},
   month = {7},
   publisher = {ACMPUB27New York, NY},
   title = {{Removal of SAT-Hard Instances in Logic Obfuscation Through Inference of Functionality}},
   volume = {29},
   year = {2024}
}

@article{Lee2015,
   author = {Yu Wei Lee and Nur A. Touba},
   doi = {10.1109/LATW.2015.7102410},
   isbn = {9781467367103},
   journal = {2015 16th Latin-American Test Symposium, LATS 2015},
   month = {5},
   publisher = {Institute of Electrical and Electronics Engineers Inc.},
   title = {{Improving Logic Obfuscation via Logic Cone Analysis}},
   year = {2015}
}

@article{Rajendran2012,
   author = {Jeyavijayan Rajendran and Youngok Pino and Ozgur Sinanoglu and Ramesh Karri},
   doi = {10.1109/date.2012.6176634},
   isbn = {9783981080186},
   issn = {15301591},
   journal = {Proceedings -Design, Automation and Test in Europe, DATE},
   pages = {953-958},
   publisher = {Institute of Electrical and Electronics Engineers Inc.},
   title = {{Logic Encryption: A Fault Analysis Perspective}},
   year = {2012}
}

@misc{niu2025flowmodularizedagenticworkflow,
      title={{Flow: Modularized Agentic Workflow Automation}}, 
      author={Boye Niu and Yiliao Song and Kai Lian and Yifan Shen and Yu Yao and Kun Zhang and Tongliang Liu},
      year={2025},
      eprint={2501.07834},
      archivePrefix={arXiv},
      primaryClass={cs.AI},
 
}

@article{Xiong2025SelfOrganizingAN,
  title={{Self-Organizing Agent Network for LLM-based Workflow Automation}},
  author={Yiming Xiong and Jian Wang and Bing Li and Yuhan Zhu and Yuqi Zhao},
  journal={ArXiv},
  year={2025},
  volume={abs/2508.13732},
}

@article{Chang2025,
   author = {Edward Y. Chang and Longling Geng},
   doi = {10.14778/3750601.3750611},
   issn = {21508097},
   issue = {12},
   journal = {Proceedings of the VLDB Endowment},
   month = {8},
   pages = {4874-4886},
   publisher = {VLDB Endowment},
   title = {{SagaLLM: Context Management, Validation, and Transaction Guarantees for Multi-Agent LLM Planning}},
   volume = {18},
   year = {2025}
}

@article{Li2024,
   author = {Zelong Li and Wenyue Hua and Hao Wang and He Zhu and Yongfeng Zhang},
   doi = {10.48550/ARXIV.2402.00798},
   journal = {arXiv.org},
   title = {{Formal-LLM: Integrating Formal Language and Natural Language for Controllable LLM-based Agents}},
   year = {2024}
}

@article{Gundawar2024,
   author = {Atharva Gundawar and Karthik Valmeekam and Mudit Verma and Subbarao Kambhampati},
   month = {11},
   title = {{Robust Planning with Compound LLM Architectures: An LLM-Modulo Approach}},
   year = {2024}
}

@article{Ayalasomayajula2024,
   author = {Avinash Ayalasomayajula and Rui Guo and Jingbo Zhou and Sujan Kumar Saha and Farimah Farahmandi},
   doi = {10.1109/mlcad62225.2024.10740198},
   month = {11},
   pages = {1-7},
   publisher = {Institute of Electrical and Electronics Engineers (IEEE)},
   title = {{LASP: LLM Assisted Security Property Generation for SoC Verification}},
   year = {2024}
}

@article{Quddus2024,
   author = {Hafiz Abdul Quddus and Md Sanowar Hossain and Ziya Cevahir and Alexander Jesser and Md Nur Amin},
   doi = {10.30420/566438009},
   isbn = {9783800764396},
   journal = {2024 Design and Verification Conference and Exhibition Europe, DVCon Europe 2024 - Proceedings},
   pages = {57-62},
   publisher = {VDE Verlag GmbH},
   title = {{Enhanced VLSI Assertion Generation: Conforming to High-Level Specifications and Reducing LLM Hallucinations with RAG}},
   year = {2024}
}

@article{Bandi2021,
   author = {Charan Bandi and Soheil Salehi and Rakibul Hassan and Sai Manoj and Houman Homayoun and Setareh Rafatirad},
   doi = {10.1109/ICSC50631.2021.00045},
   isbn = {9781728188997},
   journal = {Proceedings - 2021 IEEE 15th International Conference on Semantic Computing, ICSC 2021},
   month = {1},
   pages = {211-214},
   publisher = {Institute of Electrical and Electronics Engineers Inc.},
   title = {{Ontology-Driven Framework for Trend Analysis of Vulnerabilities and Impacts in IoT Hardware}},
   year = {2021}
}

@article{Ghimire2025_Hwrex,
   author = {Sujan Ghimire and Yu Zheng Lin and Muntasir Mamun and Muhtasim Alam Chowdhury and Farhad Alemi and Shuyu Cai and Jinduo Guo and Mingyu Zhu and Honghui Li and Banafsheh Saber Latibari and Setareh Rafatirad and Pratik Satam and Soheil Salehi},
   doi = {10.1145/3737459},
   issn = {15577309},
   issue = {6},
   journal = {ACM Transactions on Design Automation of Electronic Systems},
   month = {10},
   publisher = {Association for Computing Machinery},
   title = {{HWREx: AI-enabled Hardware Weakness and Risk Exploration and Storytelling Framework with LLM-assisted Mitigation Suggestion}},
   volume = {30},
   year = {2025}
}

@article{Hassan2023,
   author = {Rakibul Hassan and Charan Bandi and Meng Tien Tsai and Shahriar Golchin and P. D. Sai Manoj and Setareh Rafatirad and Soheil Salehi},
   doi = {10.1109/ISQED57927.2023.10129378},
   isbn = {9798350334753},
   issn = {19483295},
   journal = {Proceedings - International Symposium on Quality Electronic Design, ISQED},
   publisher = {IEEE Computer Society},
   title = {{Automated Supervised Topic Modeling Framework for Hardware Weaknesses}},
   volume = {2023-April},
   year = {2023}
}

@article{Gubbi2024,
   author = {Kevin Immanuel Gubbi and Banafsheh Saber Latibari and Muhtasim Alam Chowdhury and Afrooz Jalilzadeh and Erfan Yazdandoost Hamedani and Setareh Rafatirad and Avesta Sasan and Houman Homayoun and Soheil Salehi},
   doi = {10.1109/TCSI.2024.3364160},
   issn = {15580806},
   issue = {5},
   journal = {IEEE Transactions on Circuits and Systems I: Regular Papers},
   month = {5},
   pages = {2031-2044},
   publisher = {Institute of Electrical and Electronics Engineers Inc.},
   title = {{Optimized and Automated Secure IC Design Flow: A Defense-in-Depth Approach}},
   volume = {71},
   year = {2024}
}

@article{Lin2025,
   author = {Yu-Zheng Lin and Sujan Ghimire and Abhiram Nandimandalam and Jonah Michael Camacho and Unnati Tripathi and Rony Macwan and Sicong Shao and Setareh Rafatirad and Rozhin Yasaei and Pratik Satam and Soheil Salehi},
   month = {8},
   title = {{LLM-HyPZ: Hardware Vulnerability Discovery using an LLM-Assisted Hybrid Platform for Zero-Shot Knowledge Extraction and Refinement}},
   year = {2025}
}

@article{Asmita2025,
   author = {Asmita Asmita and Grisha Bandodkar and Sujan Ghimire and Shaurya Srivastav and Soheil Salehi and Houman Homayoun},
   doi = {10.1109/iccd65941.2025.00089},
   month = {12},
   pages = {582-589},
   publisher = {Institute of Electrical and Electronics Engineers (IEEE)},
   title = {{LLM4MCU-Onto: Leveraging LLMs for Automated Ontology Generation From Microcontroller Reference Manual}},
   year = {2025}
}

@article{Ghimire2025_Survey,
   author = {Sujan Ghimire and Muhtasim Alam Chowdhury and Banafsheh Saber Latibari and Muntasir Mamun and Jaeden Wolf Carpenter and Benjamin Tan and Hammond Pearce and Krishnendu Chakrabarty and Pratik Satam and Soheil Salehi},
   month = {6},
   title = {{Hardware Design and Security Needs Attention: From Survey to Path Forward}},
   year = {2025}
}

@inproceedings{Latibari2025,
author = {Saber Latibari, Banafsheh and Nazari, Najmeh and Sasan, Avesta and Homayoun, Houman and Satam, Pratik and Salehi, Soheil and Sayadi, Hossein},
title = {Transformers for Secure Hardware Systems: Applications, Challenges, and Outlook},
year = {2025},
isbn = {9798400714962},
publisher = {Association for Computing Machinery},
address = {New York, NY, USA},
doi = {10.1145/3716368.3735281},
booktitle = {Proceedings of the Great Lakes Symposium on VLSI 2025},
pages = {841–848},
numpages = {8},
location = {
},
series = {GLSVLSI '25}
}

@inproceedings{Veeramani2025,
author = {Pugazhenthi, Veeramani and Chowdhury, Muhtasim Alam and Ghimire, Sujan and Dharavath, Harish and Saber Latibari, Banafsheh and Salehi, Soheil},
year = {2025},
month = {08},
pages = {588-592},
title = {Power Side-Channel Leakage Assessment of Fpga-Based Spiking Neural Networks},
doi = {10.1109/MWSCAS53549.2025.11244502}
}

\end{document}